\documentclass[conference]{IEEEtran}

\usepackage{multirow}    					
\usepackage{hhline}     				 		
\usepackage{algorithm} 
\usepackage{algpseudocode}
\usepackage{enumerate}
\usepackage{wrapfig}
\usepackage{fancybox}
\usepackage{url}
\usepackage{balance}
\usepackage{amssymb,amsmath}
\usepackage{pifont}
\usepackage{xcolor}
\usepackage{tikz}
\usepackage{graphicx}
\usepackage{color}    
\usepackage{listings}
\usepackage{siunitx}
\usepackage{soul} 
\usepackage[inline]{enumitem} 

\definecolor{KWColor}{rgb}{0.37,0.08,0.25}
\definecolor{CommentColor}{rgb}{0.133,0.545,0.133}
\definecolor{StringColor}{rgb}{0,0.126,0.941}
\usepackage[utf8]{inputenc}
\usepackage[most]{tcolorbox}
\usepackage{listings}
\usepackage{upquote}
\usepackage{subfigure}
\usepackage[most]{tcolorbox}
\usepackage{enumitem}
\usepackage{color}
\usepackage{makecell}
\usepackage{booktabs}

\IEEEoverridecommandlockouts

\newcommand{\todo}[1]{}
\renewcommand{\todo}[1]{{\color{red} TODO: {#1}}}

\lstdefinestyle{JAVA}{
  language=JAVA,
  moredelim=[is][\underbar]{_}{_},
}

\newcommand{\tool}[1]{\textsc{Seeker}}

\newcommand{\mynote}[2]{\IEEEauthorrefmark{2}
    \fbox{\bfseries\sffamily\scriptsize#1}
    {\small$\blacktriangleright$\textsf{\emph{#2}}$\blacktriangleleft$}}

\definecolor{forestgreen}{HTML}{008000}

\newcommand{\kui}[1]{\mynote{kui}{\textcolor{red}{#1}}}

\title{Characterizing Sensor Leaks in Android Apps
}
\date{December 2020}

\author{
    \IEEEauthorblockN{
    Xiaoyu Sun\IEEEauthorrefmark{1}, 
    Xiao Chen\IEEEauthorrefmark{1}, 
    Kui Liu\IEEEauthorrefmark{2},
    Sheng Wen\IEEEauthorrefmark{3},
    Li Li\IEEEauthorrefmark{1}, 
    John Grundy\IEEEauthorrefmark{1}, 
    }
    \IEEEauthorblockA{\IEEEauthorrefmark{1}Monash University, Melbourne, Australia
    \\\{Xiaoyu.sun, xiao.chen, Li.Li, john.grundy\}@monash.edu
    }\thanks{Li Li is the corresponding author.}
    \IEEEauthorblockA{\IEEEauthorrefmark{2}Nanjing University of Aeronautics and Astronautics, Nanjing, China
    \\kui.liu@nuaa.edu.cn
    }
    \IEEEauthorblockA{\IEEEauthorrefmark{3}Swinburne University of Technology, Melbourne, Australia
    \\swen@swin.edu.au
    }
}

\begin{document}

\maketitle

\begin{abstract}

While extremely valuable to achieve advanced functions, mobile phone sensors can be abused by attackers to implement malicious activities in Android apps, as experimentally demonstrated by many state-of-the-art studies.
There is hence a strong need to regulate the usage of mobile sensors so as to keep them from being exploited by malicious attackers.
However, despite the fact that various efforts have been put in achieving this, i.e., detecting privacy leaks in Android apps, we have not yet found approaches to automatically detect sensor leaks in Android apps.
To fill the gap, we designed and implemented a novel prototype tool, \tool{}, that extends the famous FlowDroid tool to detect sensor-based data leaks in Android apps.
\tool{} conducts sensor-focused static taint analyses directly on the Android apps' bytecode and reports not only sensor-triggered privacy leaks but also the sensor types involved in the leaks.
Experimental results using over 40,000 real-world Android apps show that \tool{} is effective in detecting sensor leaks in Android apps, and malicious apps are more interested in leaking sensor data than benign apps.

\end{abstract}
\section{Introduction}

As of 1st January 2021, there are nearly three million Android apps available on the official Google Play app store. 
The majority of them (over 95\%) are made freely accessible to Android users and cover every aspect of users' daily life, such as supporting social networking, online shopping, banking, etc.
Many of these functionalities are supported by application interfaces provided by the Android framework, essentially fulfilled by a set of hardware-based sensors~\cite{developerandroid}.
For example, Android apps often leverage  accelerometer sensors to detect the orientation of a given smartphone and user movement, and the temperature sensor to detect the device's temperature.

Despite being needed to support the implementation of many diverse Android apps, mobile phone sensors can also be abused to achieve malicious behaviors.
There have been many reports of  apps that exploit sensors in Android devices to conduct malicious activities.
For example, Adam et al.~\cite{aviv2012practicality} have experimentally shown that the accelerometer sensor could be leveraged as a side-channel to infer mobile users' tap and gesture-based input.
Xu et al.~\cite{xu2012taplogger} have also demonstrated the possibility of this attack by presenting to the community a Trojan application named \emph{TapLogger} to silently infer user's tap inputs based on the device's embedded motion sensors.
Similarly, Schlegel et al.~\cite{schlegel2011soundcomber} have provided another Trojan application called \emph{Soundcomber} that leverages the smartphone's audio sensor to steal users' private information.

These studies have experimentally shown that the leaks of Android sensor data can  cause severe app security issues.
We argue that there is thus a strong need to invent automated approaches to detect such sensor leaks in Android apps before publishing them onto app markets.
To the best of our knowledge, existing works  focus on detecting certain types of sensor usage and its corresponding suspicious behaviors. None of them are designed as a generic approach for systematically revealing data leaks in all types of Android sensors. Also, these works mainly concentrate on discovering and understanding the usage patterns of Android embedded sensors, which do not involve completed data flow analysis to pinpoint sensitive data leaks caused by sensors.

Although many generic approaches to detect privacy leaks in Android apps have been proposed, none can be directly applied to achieve our purpose, i.e., detecting generic sensor leaks in Android apps.
Indeed, the famous FlowDroid tool has been demonstrated to be effective in detecting method-based privacy leaks in Android apps.
It performs static taint analysis on Android apps' bytecode and attempts to locate data-flow paths connecting two methods, i.e., from a \emph{source} to a \emph{sink} method.
Here, \emph{source} refers to such methods that obtain and return sensitive information from the Android framework (e.g., get device id), while \emph{sink} refers to such methods that perform dangerous operations such as sending data to remote servers.
FlowDroid has been designed as a generic approach.
It has provided a means for users to pre-define the needed \emph{source} and \emph{sink} methods.
Unfortunately,FlowDroid does not allow users to configure fields as \emph{sources} so as to support the detection of privacy leaks flowing from \emph{fields} to sensitive operations (i.e., \emph{sink}).
Since sensor data in Android is mostly provided via fields, FlowDroid cannot be directly applied to detect sensor leaks in Android apps.

To address this research gap, we designed and implemented a prototype tool, \tool{}, to automatically detect sensor data leaks in Android apps.
We  extend the open-source tool FlowDroid to support field-triggered sensitive data-flow analyses. Our new
\tool{} further performs a detailed static code analysis to infer the sensor types involved in the sensitive data-flows as the leaked sensor data is not directly associated with the sensor type. (we detail this challenge in Section~\ref{subsec:type}).
We then apply \tool{} to detect and characterize sensor leaks in real-world Android apps.
Based on 40,000 randomly selected Android apps, including 20,000 benign apps and 20,000 malicious apps, our experimental results show that \tool{} is effective in detecting sensor leaks in Android apps.
We also find that malware is more interested in obtaining and leaking sensor data than benign apps, and Accelerometer and Magnetic are among the most targeted sensors by those malicious apps.

We make the following main contributions in this work:
\begin{itemize}[leftmargin=*]
\item We have designed and implemented a prototype tool, \tool{} (\underline{Se}nsor l\underline{e}a\underline{k} find\underline{er}), that leverages static analysis to automatically detect privacy leaks originated from Android sensors.

\item We apply \tool{} to analyze both malware and benign apps at a large scale. Our results show many sensor leaks that are overlooked by the state-of-the-art static analysis tool.

\item We have demonstrated the effectiveness of our tool by evaluating the sensor leaks it highlights.

\end{itemize}

\section{Background and Motivation}
\subsection{How sensors work in Android platforms}

\begin{table*}[!h]
\small
\centering
\caption{Sensor types supported by the Android platform.} 
\vspace{-2mm}
\label{tab:sensor_types}
\resizebox{\linewidth}{!}{
\begin{tabular}{l l l } 
\hline
Sensor Type & Sensor Category  & Description \\
 \hline
 Gravity & Motion sensor & Provides a three dimensional vector indicating the direction and magnitude of gravity\\
 Linear acceleration & Motion sensor & Provides a three-dimensional vector representing acceleration along each device axis \\
 Rotation vector & Motion sensor & Provides the orientation of the device \\
 Significant motion & Motion sensor & Triggers an event each time significant motion is detected and then it disables itself \\
 Step counter & Motion sensor & Provides the number of steps taken by the user since the last reboot\\
 Step detector & Motion sensor & Triggers an event each time the user takes a step  \\
 Accelerometer & Motion sensor & Measures the acceleration applied to the device, including the force of gravity  \\
 Gyroscope & Motion sensor & measures the rate of rotation in rad/s around a device's x, y, and z axis  \\
 
 Game rotation & Position sensor & Identical to the Rotation vector sensor, except it does not use the geomagnetic field \\
 Geomagnetic rotation & Position sensor & Similar to the rotation vector sensor, but it doesn't use the gyroscope \\
 Geomagnetic field & Position sensor & Monitor changes in the earth's magnetic field \\
 Uncalibrated magnetometer & Position sensor & Similar to the geomagnetic field sensor, except that no hard iron calibration is applied\\
 Proximity sensor & Position sensor & Determine how far away an object is from a device \\
 
 Light & Environment sensor & Provides Illuminance \\
 Pressure & Environment sensor & Provides ambient air pressure \\
 Temperature & Environment sensor & Provides device temperature \\
 Ambient temperature & Environment sensor & Provides ambient air temperature \\
 Humidity & Environment sensor & Provides ambient relative humidity \\
\hline
\end{tabular} 
}
\vspace{-4mm}
\end{table*}

\begin{figure}[h!]
    \centering
    \vspace{-5mm}
    \subfigure[]{\label{fig:layers_sensor_stack}\includegraphics[width=0.48\linewidth]{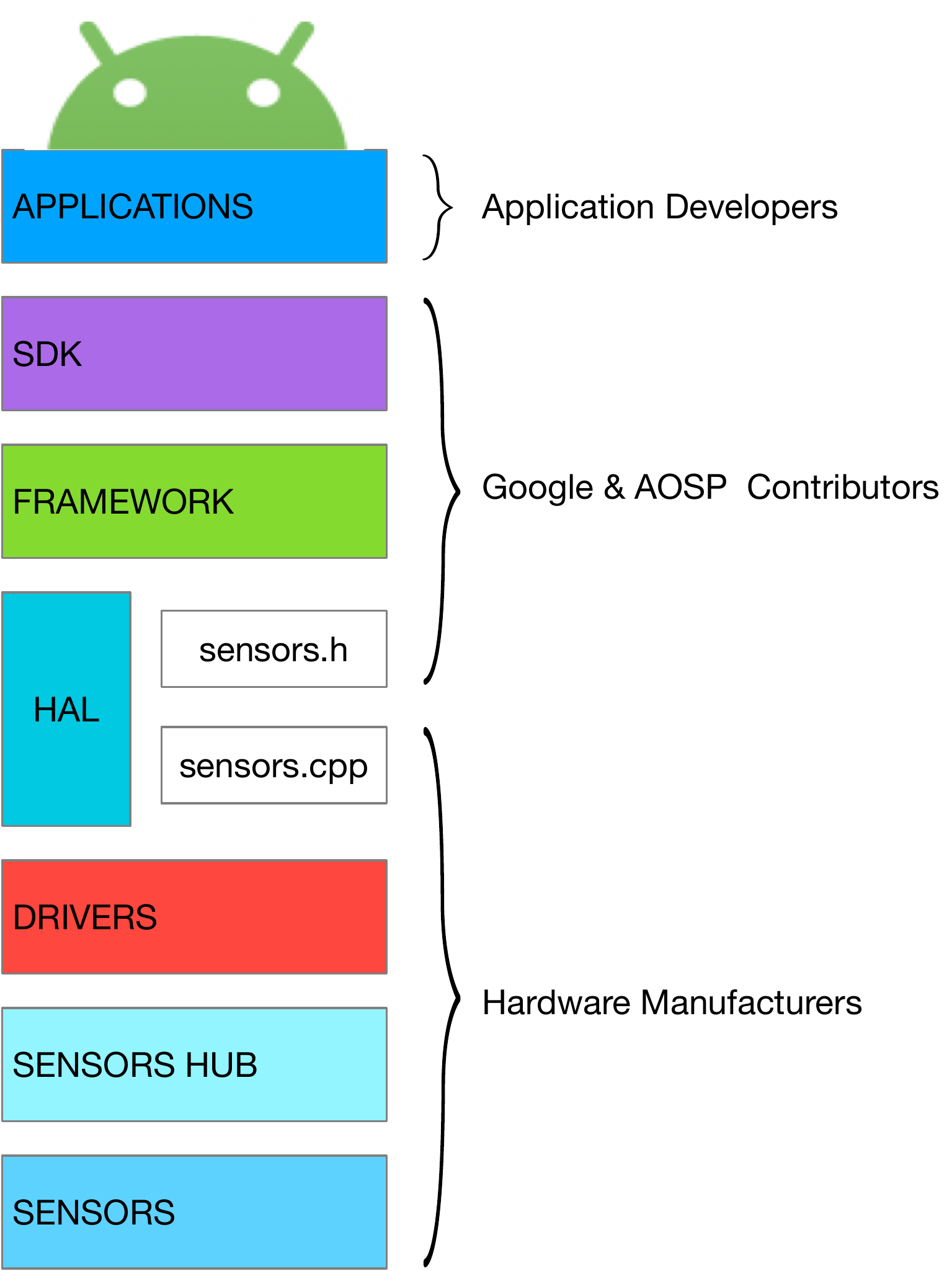}}
    \subfigure[]{\label{fig:coordinate}\includegraphics[width=0.48\linewidth]{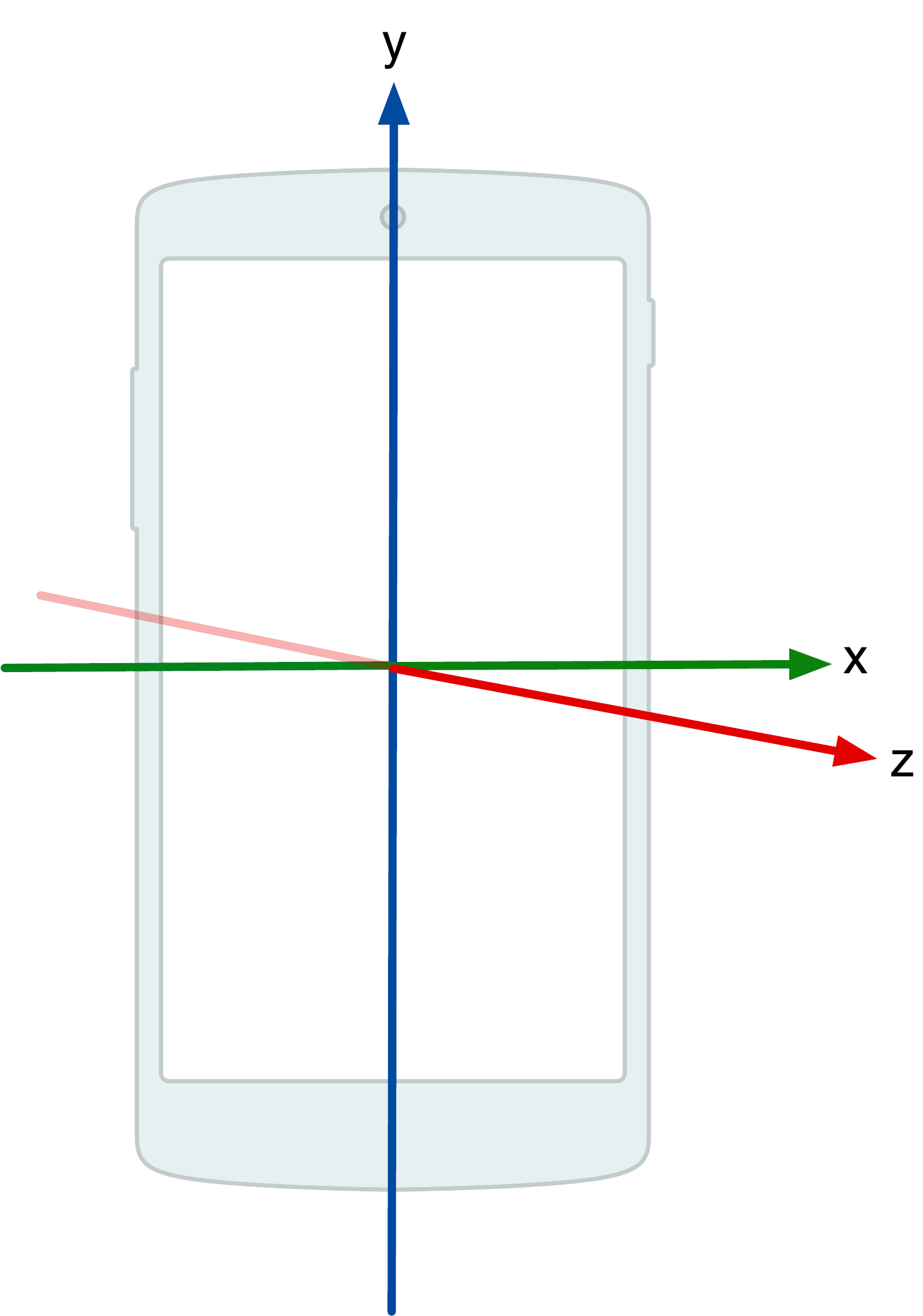}}
	\caption{Layers and Coordinate system of the Android sensor stack. Source: https://developer.android.com/guide/topics/sensors/sensors\_overview}
    \label{fig:Android_Sensor}
    \vspace{-1mm}
\end{figure}

Figure~\ref{fig:layers_sensor_stack} depicts the Android sensor stack. Sensors are Microelectromechanical systems (MEMS) chips that detect events or changes in surrounding environment. After the sensors capture the events, data is optionally passed on to the Sensors Hub. This Sensors Hub performs low-level computation as a support to the sensors, such as step counting and sensor fusion. Then the Drivers and Hardware Abstraction Layer (HAL) handles the interaction between the hardware and the Android framework. Finally, the Android apps access the sensor data through APIs provided by the Android Software Development Kit (SDK).  

In general, Android platform provides three broad categories of sensors for measuring motion, orientation, and various environmental conditions of the device:

\begin{itemize}
\item \textbf{Motion sensors}: These are used to monitor device movement, such as tilt, shake, rotation, or swing. The movement usually reflects direct user input or the physical environment around the device. Motion sensors include the accelerometer, the gyroscope, the step counter, etc.

\item \textbf{Position sensors}: These determine the physical position of a device in the world's frame of reference or the orientation of a device. Position sensors include the geomagnetic field sensor, the proximity sensor, etc.

\item \textbf{Environmental sensors}: These monitor various environmental properties, such as relative ambient humidity, illuminance, ambient pressure, and ambient temperature near the device. Examples of environmental sensors include the light sensor, the pressure sensor, etc. 
\end{itemize}

Android uses a standard 3-axis coordinate system to represent data values, as shown in Figure~\ref{fig:coordinate}. The X-axis is defined relative as horizontal, the Y-axis is vertical, and the Z-axis points towards the outside of the screen face. This coordinate system is unalterable when the device's screen orientation changes, which means the sensor's coordinate system remains the same even if the device is on the move. 

Table~\ref{tab:sensor_types} summarises the main embedded sensors supported by Android with their categories, types, and descriptions. The Android sensor framework provides both hardware-based and software-based sensors. Hardware-based sensors are accessed by reading the data directly from physical components built in the device, such as acceleration, geomagnetic field strength, or angular change. Software-based sensors derive their data from one or more of the hardware-based sensors. Examples of software-based sensors includes the linear acceleration sensor and the gravity sensor. 


\begin{lstlisting}[
caption={Example of demonstrating how to access the device's sensors.},
label=code:example_sensor_usage,
firstnumber=1]
public class SensorActivity extends Activity implements SensorEventListener {
 private SensorManager sensorManager;
 private Sensor pressure;
 private List<Sensor> deviceSensors;
 @Override
 public final void onCreate(Bundle savedInstanceState) {
  super.onCreate(savedInstanceState);
  setContentView(R.layout.main);
  // Get an instance of the sensor service, and use that to get an instance of a particular sensor.
  sensorManager = (SensorManager) getSystemService(Context.SENSOR_SERVICE);
  deviceSensors = sensorManager.getSensorList(Sensor.TYPE_ALL);
  pressure = sensorManager.getDefaultSensor(Sensor.TYPE_PRESSURE);
 }
 @Override
 public final void onAccuracyChanged(Sensor sensor, int accuracy) {
  // Do something here if sensor accuracy changes.
 }
 @Override
 public final void onSensorChanged(SensorEvent event) {
  float millibarsOfPressure = event.values[0];
  // Do something with this sensor data.
 }
 @Override
 protected void onResume() {
 //Register a listener for the sensor.
 super.onResume();
 sensorManager.registerListener(this, pressure, SensorManager.SENSOR_DELAY_NORMAL);
 }
 @Override
 protected void onPause() {
  //Unregister the sensor when the activity pauses.
  super.onPause();
  sensorManager.unregisterListener(this);}}
\end{lstlisting}

The Android sensor framework provides several APIs for developers to access its sensors and acquire raw data. We present an example in Listing~\ref{code:example_sensor_usage} to elaborate on how one identifies and determines sensor capabilities. First, to identify the sensors on a device, developers need to obtain the sensor service by calling the \texttt{getSystemService()} method and then passing the constant "Context.SENSOR\_SERVICE" as an argument (line 10). After that, developers can get a list of all sensors on a device through invoking \texttt{getSensorList(int type)}(line 11). To access a specific sensor, method \texttt{getDefaultSensor(int type)} can be called with a specific type constant (line 12).
To monitor sensor events,  the developer should implement two callback methods that are exposed through \texttt{SensorEventListener} interface, which are \texttt{onAccuracyChanged()} and \texttt{onSensorChanged()} (lines 15-17 and 19-22, respectively). Whenever a sensor detects a change, the Android system will call these two methods to report the following details to users:

 \textbf{Sensor accuracy changes}
When the sensor's accuracy changes, \texttt{onAccuracyChanged()} will provide users with a reference of the \texttt{Sensor} object and the new accuracy status of this sensor.

 \textbf{Sensor value changes}
When a sensor obtains a new value, \texttt{onSensorChanged()} will provide users with a \texttt{SensorEvent} object, which contains the accuracy of the data, the sensor object, the timestamp when the data was generated, and the new data that the sensor recorded. 


Last, the \texttt{onResume()} (lines 24-28) and \texttt{onPause()} (lines 30-34) callback methods are used to register and unregister the listener for the sensor. When an activity is paused, the related sensors should be disabled to avoid battery draining.

\begin{table}[!h]
\small
\centering
\caption{Examples of Sensor-based Cybersecurity attacks.} 
\vspace{-2mm}
\label{tab:sensor_attacks}
\resizebox{\linewidth}{!}{
\begin{tabular}{l l l } 
\hline
 Sensor Category &  Sensor Type & Attack Description \\
 \hline
 \multirow{23}{*}{Motion sensor} &
 Accelerometer & sniffing smartwatch passwords \cite{lu2018snoopy} \\
 & Accelerometer, Gyroscope & Text Inference \cite{hodges2018reconstructing}\\
 & Accelerometer, Gyroscope & Motion-based keystroke inference \cite{cai2012practicality} \\
 & Accelerometer, Gyroscope & Keystroke inference on Android \cite{al2013keystrokes} \\
 & Accelerometer & Accelerometer side channel attack \cite{aviv2012practicality} \\
 & Accelerometer & Touchscreen area identification \cite{owusu2012accessory} \\
 & Accelerometer & Decoding vibrations from nearby keyboards  \cite{marquardt2011sp}  \\
 & Gyroscope & Single-stroke language-agnostic keylogging \cite{narain2014single} \\
 & Accelerometer, Gyroscope & Inferring Keystrokes on Touch Screen \cite{cai2011touchlogger} \\
 & Accelerometer, Gyroscope & Inferring user inputs on smartphone touchscreens  \cite{xu2012taplogger}\\
 & Accelerometer, Gyroscope & Keystroke Inference \cite{bo2019know} \\
 & Accelerometer & keystrokes Inference in a virtual environment.  \cite{ling2019know}\\
 & Accelerometer, Gyroscope & Risk Assessment of motion sensor \cite{huang2019risk}\\
 & Accelerometer, Gyroscope & Infer tapped and traced user input \cite{nguyen2015using} \\
 & Accelerometer, Gyroscope & Motion-based side-channel attack  \cite{lin2019motion} \\
 & Accelerometer & Keystroke inference with smartwatch \cite{liu2015good}\\
 & Accelerometer & Motion leaks through smartwatch sensors \cite{wang2015mole}\\
 & Accelerometer & Side-channel inference attacks \cite{maiti2018side} \cite{maiti2015smart}\\
 & Accelerometer & Smartphone PINs prediction  \cite{sarkisyan2015wristsnoop} \\
 & Gyroscope & Inferring Mechanical Lock Combinations \cite{maiti2018towards} \\
 & Accelerometer, Gyroscope & Inference of private information \cite{maiti2018towards} \\
 & Accelerometer, Gyroscope & Typing privacy leaks via side-Channel from smart watch \cite{liu2019aleak}\\
 & Accelerometer, Magnetometer & Input extraction via motion sensor \cite{shen2015input} \\
 & Gyroscope & Recognizing speech \cite{michalevsky2014gyrophone} \\
 \hline
 \multirow{2}{*}{Position sensor} 
 & Magnetic & Compromising electromagnetic emanations  \cite{vuagnoux2009compromising}\\
 & Magnetic & My Smartphone Knows What You Print \cite{song2016my} \\
 & Magnetic & Location detection  \cite{block2018my} \\
\hline
\multirow{1}{*}{Environment sensor} 
 &Light Sensor &  Optical eavesdropping on displays  \cite{chakraborty2017lightspy}\\
\hline
\end{tabular} 
}
\vspace{-4mm}
\end{table}

\subsection{Motivation}
\label{subsec:motivation}

Sensors have been widely adopted for launching side-channel attacks against smart devices \cite{sikder2021survey}.   
Table ~\ref{tab:sensor_attacks} summarizes a diverse set of sensor-based attacks targeting smartphones and smartwatches. Since accessing sensitive sensor data does not require any security checks (e.g., permission check), attackers can easily trigger malicious behaviors by making use of such data. As revealed in the table, generally, sensor leakage are performed with the aim of (1) keystroke inference, (2) task inference (refers to a type of attack which reveals the information of an on-going task or an application in a smart device), (3) location inference, and (4) eavesdropping. For example, motion and position sensors can be exploited for keystroke inference, leading to severe privacy leaks such as passwords, credit card information, etc. 
Light sensor is found to eavesdrop acoustic signals in the vicinity of the device, causing private information leak. Magnetic sensors can be exploited to compromise electromagnetic emanations, which would affect the confidentiality of the devices.

\begin{figure}[!t]
    \centering 
    \includegraphics[width=0.45\textwidth]{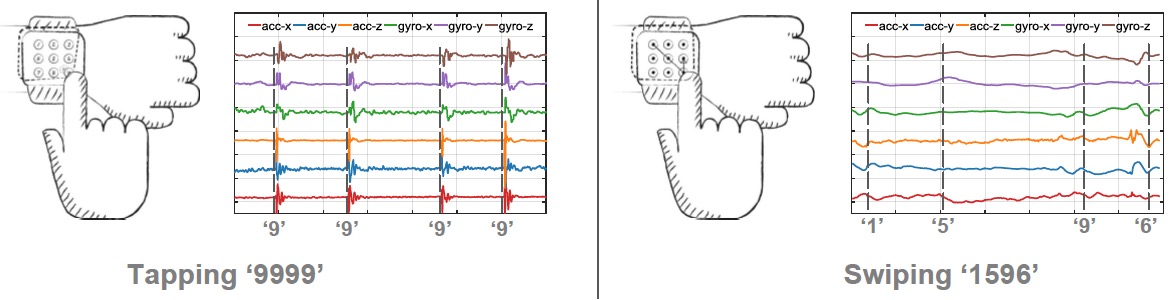}
    \vspace{-1mm}
    \caption{The snoopy example of sniffing smartwatch passwords via censoring motion sensor data \cite{lu2018snoopy}.}
    \vspace{-3mm}
    \label{fig:snoopy_example}
\end{figure}

As a concrete example, Lu et al. \cite{lu2018snoopy} revealed that sensitive intercepting password could be accessed through motion data on the smartwatch's onboard sensors. They proposed \emph{Snoopy}, a password extraction and inference approach via sensor data for PIN attack, which could affect smartwatch users in a non-invasive way.
\emph{Snoopy} extracts the segments of motion data when users entered passwords and then applies deep learning techniques to infer the actual passwords. Figure~\ref{fig:snoopy_example} gives two examples of the differences of the motion sensor data changes when the user swipes or taps a password on a smartwatch.
\emph{Snoopy} demonstrates the feasibility of sensor data leaks by intercepting password information entered on smartwatches. 
Such real-world sensor-enabled attacks motivated us to provide automatic tools for characterizing universal sensor leaks in Android Apps that have been long overlooked.

\begin{figure*}[!h]
    \centering
    \includegraphics[width=0.7\linewidth]{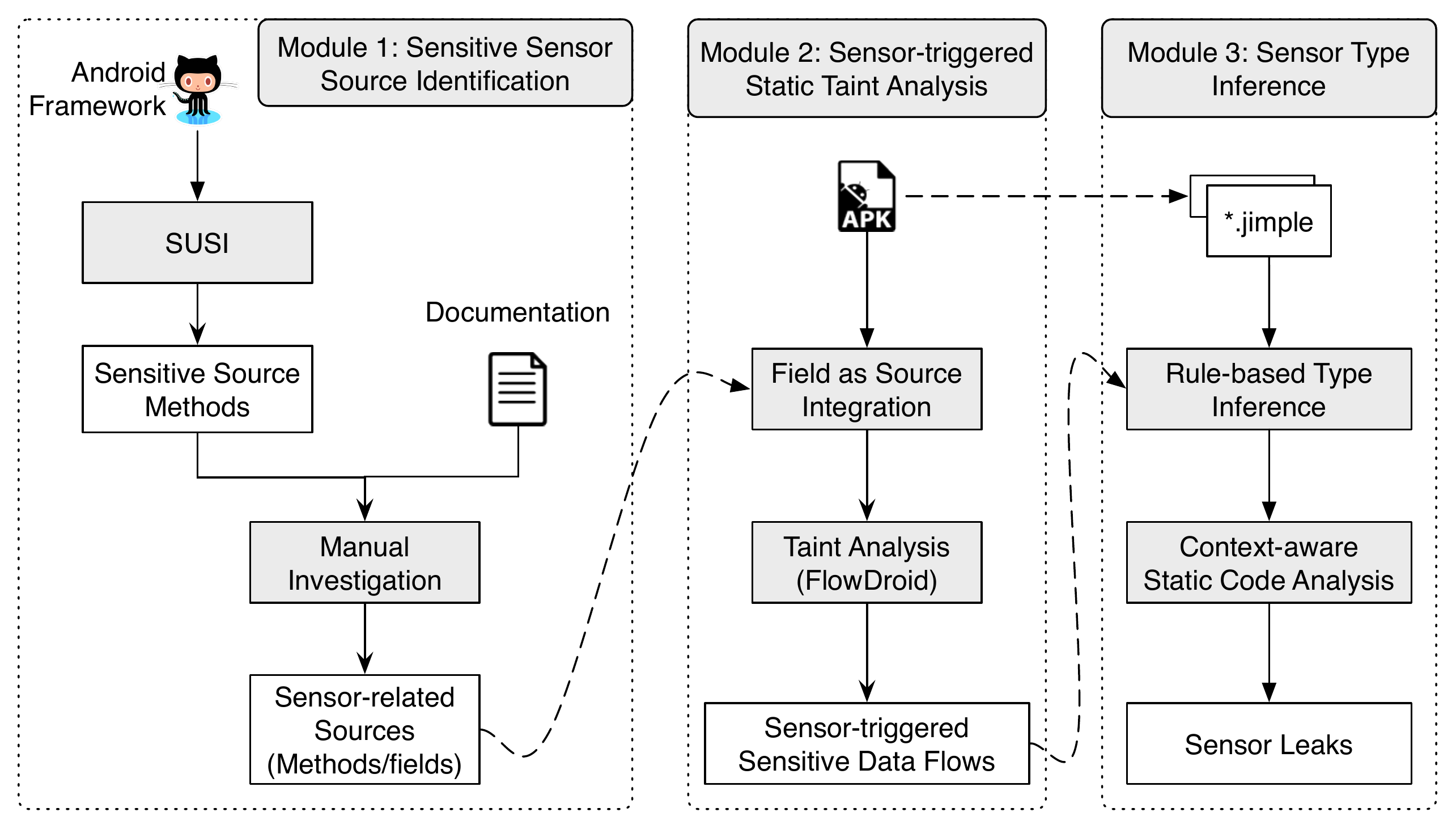}
    \vspace{-3mm}
	\caption{The working process of our approach.}
    \label{fig:methodology}
\end{figure*}

\vspace{-1mm}
\section{Approach}

This work aims to automatically detect information leaks of onboard sensors in Android apps. To this end, we design and implement a prototype tool called \tool{} for achieving this purpose. Figure \ref{fig:methodology} describes the overall working process of \tool{}, which is mainly made up of three modules, namely Sensitive Sensor Source Identification, Sensor-triggered Static Taint Analysis and Sensor Type Inference.

\vspace{-1mm}
\subsection{Sensitive Sensor Source Identification}
\label{subsec:source}
The first module, \emph{Sensitive Sensor Source Identification}, aims to identify sensor-related sources that access and obtain sensitive information related to the device's sensors. As reported by Liu et al.~\cite{liu2018discovering}, Android sensor data can be obtained through invoking sensor-related APIs or directly accessing local fields in which the sensor data is stored.
In this work, we take both of these types into consideration, aiming at pinpointing all the possible sensor-triggered privacy leaks.

To do this we need to identify all the sensor-related sources, including both Android methods and fields.
For Android methods, we use the well known SUSI tool~\cite{arzt2013susi} to obtain sensor-related source methods. 
SUSI is a novel machine-learning guided approach that scans Android API's source code to predict \emph{source} and \emph{sink} methods, based on a training set of hand-annotated sources and sinks.
In this work, we launch SUSI on the latest Android Open Source Project (i.e., AOSP version 11.0) and manually filter out non-sensor related source methods.

To identify sensor-related fields (as sources), there is no existing approach to achieve such a purpose. We resort to a manual process of going through the Android Developers' Documentation to identify source fields storing sensitive sensor information. The identified fields are then discussed and confirmed by the authors by measuring whether leaking such information would potentially expand the attack surface to users' privacy. Finally, we identified 79 fields and 20 methods as the sources. Table \ref{tab:sensor_sources} lists the selected sources that indeed introduce leaks in our experimental dataset. A full list of field and method sources can be found in the \emph{SourcesAndSinks.txt} file of our open-source project\footnote{https://github.com/MobileSE/SEEKER}.

\begin{table}[!h]
\small
\centering
\caption{The list of sensitive sensor sources.}\vspace{-1mm}
\label{tab:sensor_sources}
{
\begin{tabular}{l c c} 
\hline
Sensor-related Source & Source Type \\
\hline
SensorEvent\#values & Field  \\
SensorEvent\#timestamp & Field \\
Sensor\#getName() & Method \\
Sensor\#getVendor() & Method  \\
Sensor\#getVersion() & Method  \\
SensorManager\#getDefaultSensor(int) & Method  \\
Sensor\#getMaximumRange() & Method  \\
SensorManager\#getSensorList(int) & Method  \\
Sensor\#getType() & Method  \\
Sensor\#getResolution() & Method  \\
Sensor\#getPower() & Method  \\

 \hline

\hline
\end{tabular} 
}
\vspace{-2mm}
\end{table}

\vspace{-1mm}
\subsection{Sensor-triggered Static Taint Analysis}
The ultimate goal of \tool{} is to detect sensor-related data leaks. To this end, we implement the \emph{Sensor-triggered Static Taint Analysis} module that extends state-of-the-art tool FlowDroid \cite{arzt2014flowdroid} to facilitate sensor-related data leak detection. FlowDroid detects data leaks by computing data flows between sources and sinks. FlowDroid defined a sensitive data flow happens when a suspicious “tainted” information passes from a source API (e.g., \texttt{getDeviceId}) to a sink API (e.g., \texttt{sendTextMessage}). 

FlowDroid is a state-of-the-art tool and it provides a highly precise static taint-analysis model, especially for Android applications. However, FlowDroid only takes API statements as sources or sinks, leading to false negatives because of the lack of field-triggered sources. Thus, in this work, we extend FlowDroid by supporting field statement as sources, so as to pinpoint data leaks originated from specific field sources of interest. 

Our preliminary study discovered that certain sensor-related data leaks are sourced from data stored in class fields (e.g., android.hardware.SensorEvent\#values). We therefore implemented our own class that implements the \texttt{ISourceSinkDefinitionProvider} interface in FlowDroid for supporting the declaration of fields as sources. Also, based on the feature of class fields, we defined a new model names AndroidField extends from \texttt{SootFieldAndMethod}. After loading a specific field statement from \emph{source\&sink.txt} file, we apply a field pattern regular expression to convert it to the AndroidField model.

FlowDroid has the ability to compute data flow connections between all possible statements. In the implementation of FlowDroid, \texttt{ISourceSinkManager} interface marks all statements as possible sources and then records all taint abstractions that are passed into \texttt{getSourceInfo()}. To that end, we pass the constructed field model as a source statement to the following taint analysis process. In this way, sensitive data flow can be detected starting at given field source statements.

\subsection{Sensor Type Inference}
\label{subsec:type}
The primary goal of \tool{} is to detect data leaks from Android platform sensors.
With the help of FlowDroid's taint analysis, \tool{}'s second module can detect sensor-triggered sensitive data flows.
Unfortunately for the field-triggered ones, the identified data-flows only show that there is sensor data leaked but do not tell from which sensor the data is collected.
The sensor type information is important for helping security analysts understand the sensor leaks.
Therefore, in our last module, we identify the types of sensors that are leaking information.

To identify which sensors exist on a specific Android device, we first get a reference to the sensor service by creating an instance of the \texttt{SensorManager} class via calling the \texttt{getSystemService()} method with \emph{SENSOR\_SERVICE} argument. After that, we can determine available sensors on the device by calling the \texttt{getSensorList()} method. 
The \texttt{getSensorList()} method returns a list of all available sensors on the device by specifying constant \emph{TYPE\_ALL} as the parameter. A list of all sensors from a given type can also be retrieved by replacing the parameter as the constants defined for corresponding sensor types, such as \emph{TYPE\_GYROSCOPE, TYPE\_LINEAR\_ACCELERATION}, etc. We can also determine whether a specific type of sensor exists by calling the \texttt{getDefaultSensor()} method with the target type constant (the same as the ones passed in to \texttt{getSensorList()} method). If a device has that type of sensor, it will return an object of that sensor. Otherwise, null will be returned.

We use a rule-based strategy to identify the sensor type of a leak in the case of only one sensor registered in the given app. To do this, \tool{} obtains the sensor type by looking into the type constant in the \texttt{getDefaultSensor()} statement. For instance, \texttt{getDefaultSensor(Sensor.TYPE \_ACCELEROMETER)} indicates that the Accelerometer sensor is obtained. We can then reasonably assume that all sensor-related data leaks in the class are associated with the identified sensor (because only this sensor is registered).

\begin{lstlisting}[
caption={An example of sensor type usage with switch branch.},
label=code:example_sensor_type_usage,
firstnumber=1]
public class MainActivity extends AppCompatActivity implements SensorEventListener{
@Override
public void onSensorChanged(SensorEvent sensorEvent) {
 switch(sensorEvent.sensor.getType()) {
  case Sensor.TYPE_ACCELEROMETER:
   accX = sensorEvent.values[0];
   accY = sensorEvent.values[1];
   accZ = sensorEvent.values[2];
   ...
  case Sensor.TYPE_GYROSCOPE:
   gyroX = sensorEvent.values[0] * 5;
   gyroY = sensorEvent.values[1] * 5;
   gyroZ = sensorEvent.values[2] * 5;
   ...
  case Sensor.TYPE_ROTATION_VECTOR:
   rvX = sensorEvent.values[0];
   rvY = sensorEvent.values[1];
   rvZ = sensorEvent.values[2];
   ...
}}}
\end{lstlisting}

In the case of multiple sensors registered in the given app, we further leverage context-aware static code analysis to find the connection between sensor types and the leaked field data. Firstly, we locate the invocation statement of API \texttt{android.hardware.SensorManager\#getDefault Sensor(int)} in the \texttt{onSensorChanged()} method. In the multiple sensors scenario, different sensor's behavior is handled in a conditional branch (e.g. if-then-else statement or switch statement). We then apply context-aware static code analysis to detect the code branch that contains the taint sensor source statement, based on which we then resolve the sensor type in the branch condition.

We further elaborate on the context-aware static code analysis with an example presented in Listing~\ref{code:example_sensor_type_usage}. The code snippet in the Listing shows an example of how multiple sensors are handled with \texttt{onSensorChanged(android .hardware.SensorEvent)} method.  Android determines the activated sensor by matching the \texttt{sensorEvent.sens or.getType()} method (line 4). For example, if \texttt{get Type()} returns \texttt{Sensor.TYPE\_ACCELEROMETER} (line 5), the data obtained by \texttt{sensorEvent.values} is associated with the Accelerometer sensor (lines 6-8); if \texttt{getType()} returns \texttt{Sensor.TYPE\_GYROSCOPE} (line 10), the data contained in \texttt{sensorEvent.values} is accordingly associated with the current activated sensor, i.e., Gyroscope (lines 11-13). 

\section{Experimental Setup and Results}

\tool{} is designed to expose the data leak issues of sensors in Android apps.
We investigate the feasibility and effectiveness of detecting sensor leaks in Android apps with the following three research questions: 

\begin{itemize}
   \item{\bf RQ1:} {\em Can \tool{} effectively detect sensor leaks in Android apps?} This research question aims to investigate the feasibility of detecting sensor leaks in Android apps with \tool{}.

    \item {\bf RQ2:} {\em To what extent diverse sensor leaks can be identified by \tool{}?} With this research question, we explore the sensor types related to the identified sensitive data leaks, and investigate to what extent such sensor leaks are targeted by attackers. 

    \item {\bf RQ3:} {\em Is \tool{} efficient to detect the sensor leaks in Android apps?} In this study, we leverage the time costs of detecting sensor leaks to assess the efficiency of \tool{}.
\end{itemize}

\subsection{Experimental Setup}

To answer the aforementioned research questions, we build the experimental dataset with a \textit{malware} set and a \textit{benign} set. The \textit{malware} set contains 20,000 Android apps including malware downloaded from VirusShare repository \cite{Virusshare} that were collected between 2012 and 2020. The 20,000 Android apps in \textit{benign} set are crawled from the official Google Play store. All of the 40,000 apps are submitted to VirusTotal \cite{Virustotal}, the online scan engines aggregating over 70 anti-virus scanners (including the famous Kaspersky, McAfee, Kingsoft anti-virus engines), to check whether they contains viruses or not.
For the \textit{malware} set, we select the malware Android apps that have been labeled by at least
five anti-virus engines to ensure their maliciousness, while for the \textit{benign} set, the Android apps that are not tagged by any anti-virus engines are selected. 
\tool{} is designed to detect sensor leaks, thus we filter out the Android apps without any onboard sensors by checking whether their code contains the string ``\texttt{android.hardware.sensor}''. 
The final experimental dataset used in this study consists of 6,724 malware apps and 12,939 benign apps (cf. the 3rd column of Table~\ref{tab:sensor_leak_results}).
Our experiment runs on a Linux server with Intel(R) Core(TM) i9-9920X CPU @ 3.50GHz and 128GB RAM. The timeout setting for analyzing each app with \tool{} is set 20 minutes. 

\subsection{RQ1 -- Feasibility of Detecting Sensor Data Leaks}
\begin{figure}[!t]
    \centering
    \vspace{-1mm}
    \includegraphics[width=0.85\linewidth]{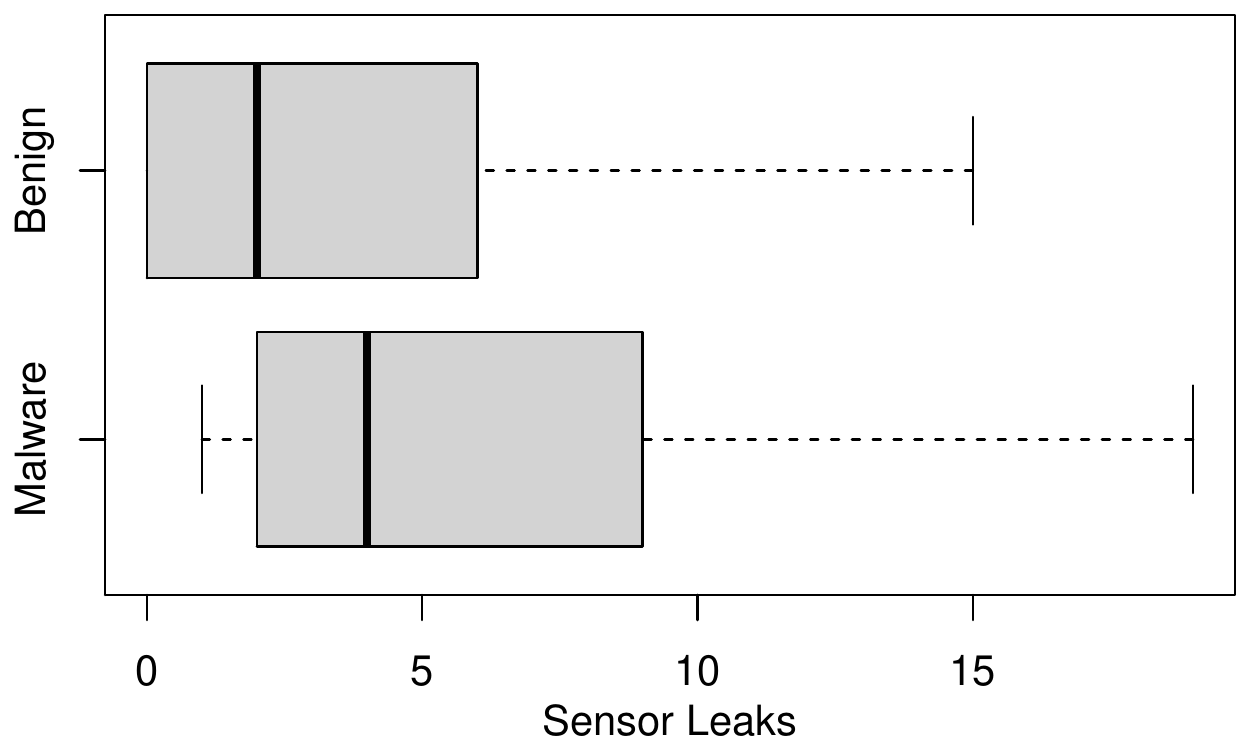}
    \vspace{-2mm}
	\caption{Distribution of sensor leaks in each app.}
    \label{fig:Android_method_field_Sensor}
    \vspace{-2mm}
\end{figure}

Our first research question evaluates the feasibility of \tool{} on detecting sensor leaks in Android apps, of which results are illustrated in Table~\ref{tab:sensor_leak_results}.
For the quantitative aspect, 9,905 potential sensor leaks are identified by \tool{} in 1,596 apps. 
On average, one Android app could be injected with six sensor leaks.
It indicates that the sensor leaks could exist in Android apps which might have been overlooked by the security analysts of Android apps.
From the malicious aspect, 14.4\% (967 out of 6,724) malware apps are identified with sensor leaks, while 4.9\% (629 out of 12,939) benign apps are identified with such leaks.
Figure~\ref{fig:Android_method_field_Sensor} further presents the number of sensor leaks detected in each Android app, which shows that each malware app could be identified with more sensor leaks than the benign one.
It is significantly confirmed by the Mann-Whitney-Wilcoxon (MWW) test \cite{fay2010wilcoxon}, of which the resulting \emph{p-value} is less than $\alpha = 0.001$. 
All of these results imply that malware apps have a higher possibility of containing sensor leaks than benign apps.


{\bf Note that:} there is lack of the ground-truth dataset about the sensor data leaks in Android apps.
To address this limitation, we consider a sensor leak existing in an Android app if there is the data flow interaction between sensor-related sources (i.e., class fields or methods) and sinks.
With this criterion, we manually checked the 229 sensor leaks detected by \tool{} in 20 randomly selected apps (10 malware apps and 10 benign apps). 
There are only 4 false-positive identified sensor leaks among the 229 identified sensor leaks in the 20 Android apps, which are caused by inaccurate data-flow analysis results of FlowDroid (we detail this limitation cased by FlowDroid in Section \ref{subsec:limitations}). 
Such results show that \tool{} is capable of identifying the sensor leaks in Android apps.
Simultaneously, it raises a major alarm for security analysts to pay attention to sensor leaks in Android apps that are not protected by the Android permission mechanism.

\begin{table}[!t]
\centering
\caption{Experimental results of the detected sensor leaks.}
\label{tab:sensor_leak_results}
{
\begin{tabular}{c c c |c c} 
\toprule
\textbf{Dataset} &
\textbf{\# apps} &
\makecell[c]{\textbf{\# selected}\\\textbf{apps}} &
\makecell[c]{\textbf{\# apps identified}\\\textbf{with sensor leaks}} &
\makecell[c]{\textbf{\# identified}\\\textbf{sensor leaks}} \\
\hline
Malware & 20,000 & 6,724 & 967 & 6,103 \\
\hline
Benign & 20,000 & 12,939 & 629 & 3,802 \\
\hline
Total & 40,000 & 19,663 & 1,596 & 9,905\\
\bottomrule
\end{tabular} }
\end{table}

\begin{tcolorbox}[title=\textbf{RQ1 \ding{43} Feasibility and Effectiveness}, left=2pt, right=2pt,top=2pt,bottom=2pt]
\tool{} is capable of automatically detecting sensor leaks in Android apps. 
Malware apps present higher possibility of committing sensor leaks than benign apps, and the sensor leaks might be ignored by security analysts.
\end{tcolorbox}

\subsection{RQ2 -- Characterization of Sensor Leaks}
\paragraph*{Sources Triggering Sensor Leaks}
The data leaks in sensor of Android apps are mainly triggered by two kinds of source: field and method (cf. Section~\ref{subsec:source}).
As presented in Table~\ref{tab:sensor_leak_results2}, $\sim$80\% (= 7941/9905) of identified sensor leaks are triggered by the method sources. For the benign Android apps, $\sim$85.8\% sensor data leaks are sourced from methods, while in the malware Android apps, $\sim$76.6\% leaks are originated from methods.
Table~\ref{tab:sensor_leak_types_method} lists the top-10 most frequent sources triggering sensor leaks identified by \tool{}, which are 8 {\tt getter} methods and 2 public fields from {\tt Sensor}-related classes. 
We observe that the most frequent leaking source is the method {\tt android.hardware.SensorManager\# getDefaultSensor(int)} that is used to get the specific sensor of a given type, which is followed by the field \texttt{values} of class \texttt{SensorEvent}.
The leaking source {\tt android.hardware.SensorManager\#getDefaultS ensor(int)} occupies $\sim$89.1\% (7074 out of 7941) of the method-triggered sensor leaks,
and the field {\tt SensorEve nt\#values} occupies $\sim$95.6\% (1877 out of 1964) of the field-triggered sensor leaks.
The sensor leaks triggered by the two sources occupy $\sim$90\% of all identified sensor leaks.

\begin{table}[!ht]
\centering
\caption{Number of identified method/field-triggered sensor leaks.}
\label{tab:sensor_leak_results2}
{
\begin{tabular}{l c c} 
\toprule
&
\textbf{\# identified method leaks} &
\textbf{\# identified field leaks} \\
\hline
Malware & 4,677 & 1,426\\
\hline
Benign  & 3,264 & 538\\
\hline
Total &   7,941 & 1,964\\
\bottomrule
\end{tabular} }
\end{table}


\begin{table}[!ht]
\centering
\vspace{-3mm}
\caption{Top-10 frequent leaking sources.}
\label{tab:sensor_leak_types_method}
\resizebox{\linewidth}{!}{
\begin{tabular}{l c c c} 
\hline
\multirow{1}{*}{\textbf{Sensor Sources}} &
\multirow{1}{*}{\textbf{Malware}} &
\multirow{1}{*}{\textbf{Benign}} &
\multirow{1}{*}{\textbf{Total}} \\
\hline
SensorManager\#getDefaultSensor(int) & 4,326 & 2,748 &  7,074\\
\hline
SensorEvent\#values & 1,358 & 519 &  1,877\\
\hline
SensorManager\#getSensorList(int) & 114 & 121  & 235 \\
\hline
Sensor\#getType() & 123 & 6 & 129 \\
\hline
Sensor\#getName() & 29 & 82 & 111 \\
\hline
Sensor\#getMaximumRange() & 19 & 73  & 92 \\
\hline
SensorEvent\#timestamp & 68 & 19 & 87 \\
\hline
Sensor\#getVendor() & 12 & 74  &  86 \\
\hline
Sensor\#getVersion() & 11 & 69  &  80 \\
\hline
Sensor\#getResolution() & 13 & 64  & 77 \\
\hline
\end{tabular} }
\end{table}



\paragraph*{Sensor Types of Field-triggered Sensor Leaks}
The sensor type is essential for deepening the understanding of sensor data leaks, i.e., knowing from which sensor the data is originally collected, as by default, this information is not given in field-triggered sensor leaks (e.g., sourced from the field variable {\tt values} in class {\tt SensorEvent}).
The last module of \tool{} is hence dedicated to infer the sensor types of such leaks.
Overall, in the 1,964 identified field-triggered sensor leaks, \tool{} successfully infers the corresponding sensor types for 1,923 (97.9\%) of them.
After manually checking the unsuccessful cases, we find that the 41 failed cases are mainly caused by the mistaken usage of sensors which can cause the sensors unexpected functional behavior, such as lacking sensor register information.
This high success rate demonstrates the effectiveness of \tool{} in pinpointing the sensor types associated with sensor data leaks.

\begin{table}[!t]
\centering
\caption{Top-10 frequent sensor types of field-triggered sensor leaks.}
\label{tab:sensor_leak_types_field}
\resizebox{\linewidth}{!}{
\begin{tabular}{l c c c} 
\hline
\multirow{1}{*}{\textbf{Sensor Type}} &
\multirow{1}{*}{\textbf{Malware}} &
\multirow{1}{*}{\textbf{Goodware}} &
\multirow{1}{*}{\textbf{Total}} \\
\hline
ACCELEROMETER & 1,068 & 304 & 1,372 \\
\hline
MAGNETIC\_FIELD & 131 & 50  & 181\\
\hline
ORIENTATION & 92 & 84 &  176\\
\hline
PROXIMITY & 12 & 32 & 44\\
\hline
LINEAR\_ACCELERATION & 40 & 4  & 44 \\
\hline
STEP\_COUNTER & 14 & 9 & 23\\
\hline
TEMPERATURE & 12 & 8 & 20 \\
\hline
GYROSCOPE & 7 & 8  & 15 \\
\hline
PRESSURE  & 1 & 13 & 14 \\ 
\hline
LIGHT & 6 & 5  & 11 \\
\hline
\end{tabular} }
\end{table}

We further investigate the true-positive rate of the successfully inferred sensor types. Due to the lack of the ground-truth dataset of related sensor types for sensor leaks, we resort to a manual inspection on the source code of 20 randomly selected apps (10 malware apps and 10 benign apps), each of which is identified with at least one field-triggered leak (86 field-triggered sensor leaks in total). 
All leaks are confirmed with true-positive inferred sensor types, which implies that \tool{} is effective in inferring the sensor types of field-triggered leaks. 


Table~\ref{tab:sensor_leak_types_field} presents the top 10 leaking sensor types of the identified field-triggered sensor leaks.
The type ``Accelerometer'' is the sensor type of 74.9\% and 56.5\% of identified field-triggered sensor leaks in the \textit{malware} apps and the \textit{benign} apps, respectively.
Android apps widely use the Accelerometer to monitor device motion states by measuring the acceleration applied to a device on three physical axes (i.e., x, y, and z axes). The motion data captured by the Accelerometer can be further processed or analyzed. For example, \emph{Smart-Its Friends} \cite{holmquist2001smart} pairs two devices by acquiring Accelerometer data in a shared wireless medium. Pirttikangas et al. \cite{pirttikangas2006feature} reported that the Accelerometer in smartphones can be used to track the accurate activity of users, such as brushing teeth and sitting while reading newspapers. Such information can also be utilized to steal the PIN of a device through side-channel attacks (such as \cite{lu2018snoopy} and \cite{giallanza2019keyboard}).

Apart from the Accelerometer, the other frequent sensor types of field-triggered sensor leaks include MAGNETIC\_ FIELD, ORIENTATION, PROXIMITY, LINEAR\_ACCEL- ERATION, STEP\_COUNTER, TEMPERATURE, GYROSCOPE, PRESSURE and LIGHT. These sensors are also likely to be used to harm the user's privacy. 
Biedermann et al. \cite{biedermann2015hard} stated that the magnetic field sensor can be exploited to detect what type of operating system is booting up and what application is being started. The orientation sensor can wiretap the device's orientation without requesting any permission, which can be used by attackers to infer the user's PIN. The proximity sensor data can be a trigger to automatically start a phone call recording when users hold the smartphone against their face to make a call. The individual step details can be stored by collecting data from the step counter sensor when the app runs in the background. Temperature, pressure and light sensors are also widely used in IoT devices to monitor environmental conditions, while the gyroscope sensor is utilized to verify the user's identity \cite{sikder2021survey}.

\paragraph*{Case Study}
Here we show two real-world apps that leaks the sensor data, which could be  leveraged by attacker to achieve malicious goals. 

\begin{lstlisting}[style=JAVA, escapechar=\%,
caption={Example of a sensor leak in {\em com.n3vgames.android.driver}.},
label=code:Case_Study_time_stamp,
firstnumber=1]
final class a.b.b implements SensorEventListener{
 public void onSensorChanged(SensorEvent var1){
    float var5 = var1.values[0];
    float var6 = var1.values[1];
    float var7 = var1.values[2];
    Log.v("WindowOrientationListenerN3V", "Raw acceleration vector: x=" + var5 + ", y=" var6 + ", z=" + var7);
}}
\end{lstlisting}

Listing \ref{code:Case_Study_time_stamp} showcases a typical sensor leak case in real-world apps. The code snippet is excerpted from a malicious app {\em com.n3vgames.android.driver}.
This app collects raw accelerometer data from the class field \texttt{SensorEven\#values[]} (lines 3-5), and then leak them through invoking \emph{android.util.Log} API (line 6). The app is flagged as a Trojan that downloads additional executable content from a remote server. While leaking such information may not direct link to its malicious behaviour, it expands the attack surface to the attackers. For example, the sensor information can be used to predict the device's motion state, which may lead to a stealthier attack (e.g., downloading malicious content when the device is not in use). It is worth noting that Zhang et al. \cite{zhang2019using} have demonstrated the possibility of using the sensor information to launch stealthy attack for taking control of an Android phone via Google's voice assistant.

\begin{figure}[t!]
    \centering
    \includegraphics[width=1\columnwidth]{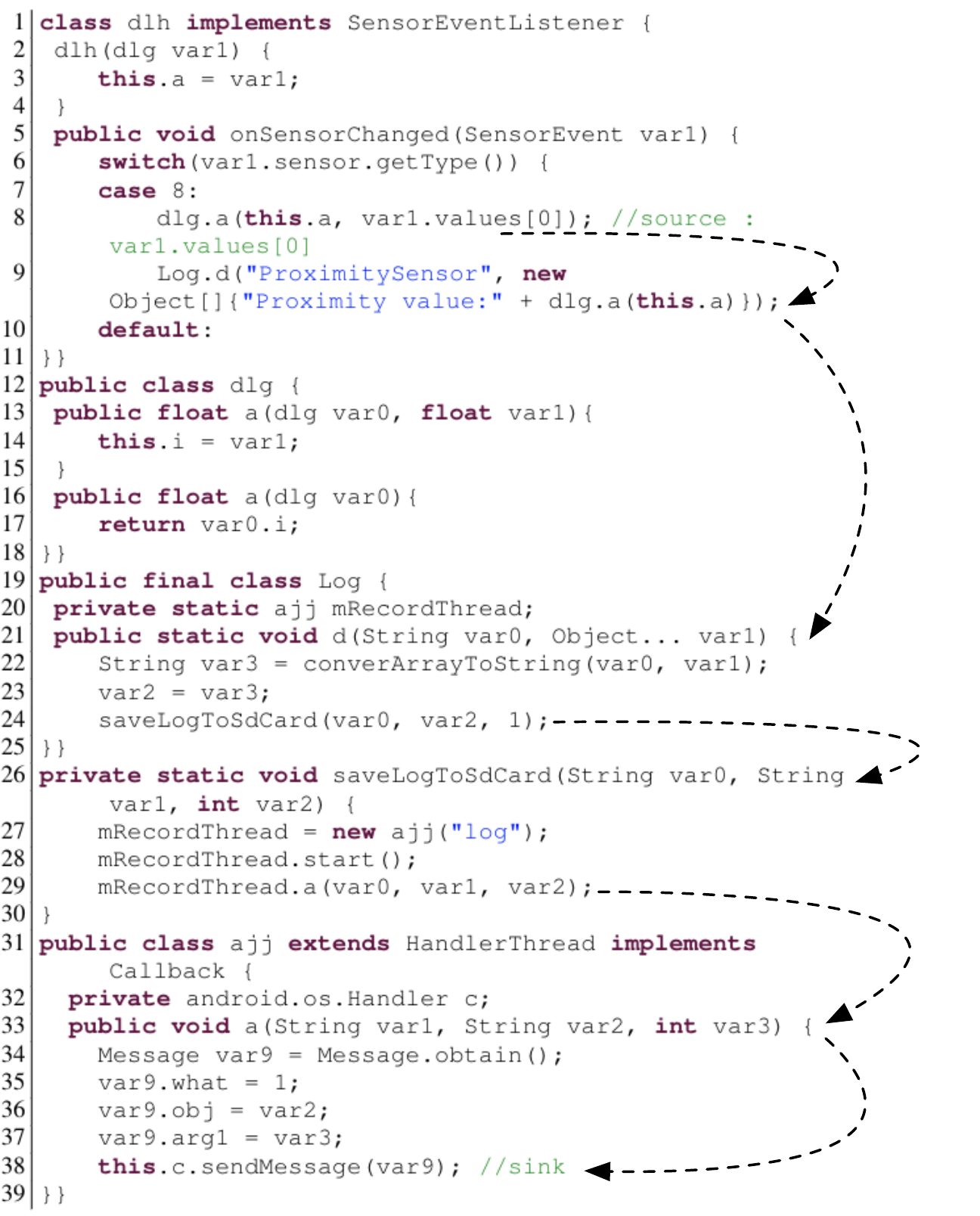}
	\caption{Code snippet of sensor value leak excerpted from \textit{com.tencent.pb}.}
    \label{fig:case_study_2}
\end{figure}

Figure \ref{fig:case_study_2} shows another example derived from a phone book app \emph{com.tencent.pb}. It collects the Proximity sensor data (line 8) and eventually send it out through {\tt sendMessage} method (line 41) in a asynchronous thread. The  Proximity sensor data was passed as a parameter of method {\tt dlg.a(dlg, float)} (line 8), then the data was stored in the class field {\tt i} of object \texttt{dlg}. The data flows through the method \emph{Log.d(String, Object...)} (line 9), which obtains the field variable \texttt{i} of object \texttt{dlg} as the second parameter via \texttt{dlg.a(dlg)} (line 23). Finally, the tainted parameter passed on to the method \texttt{Log.saveLogToSdCard(String, String, int)}, which creates a new thread (line 29-33) and send the sensor data out (line 35-42).

\begin{tcolorbox}[title=\textbf{RQ2 \ding{43} Characterizing Sensor Leaks}, left=2pt, right=2pt,top=2pt,bottom=2pt]
\tool{} is capable of inferring sensor types and pinpointing the corresponding  source sensors for data leaks. Our results show that the Accelerometer leaks the most sensor data, both in malware samples and benign apps. The most leaking sources are the method \textit{SensorManager\#getDefaultSensor(int)} and the field \textit{SensorEvent\#values}, the latter of which has been frequently leveraged in malicious behaviours such as inferring user's PIN.
\end{tcolorbox}

\begin{figure}[!h]
    \centering
    \includegraphics[width=0.85\linewidth]{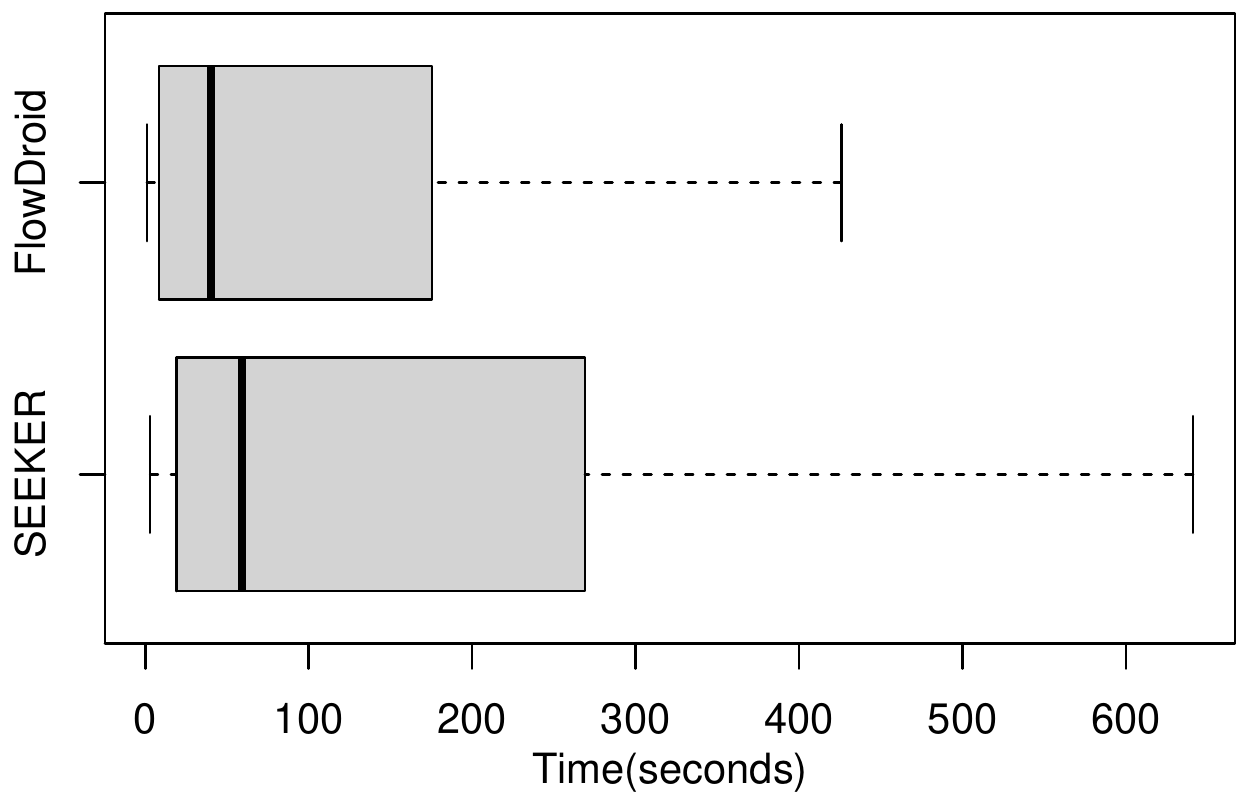}
    \vspace{-2mm}
	\caption{Distribution of time Performance spent to analyze an app by FlowDroid and \tool{}, respectively.}
    \label{fig:time_performance}
    \vspace{-3mm}
\end{figure}

\subsection{RQ3 -- Runtime Overhead}
\tool{} extends FlowDroid to detect sensor-related data leaks and for inferring the sensor types involved in the leak. We evaluate the runtime overhead of \tool{} and compare it with the original FlowDroid. Figure \ref{fig:time_performance} shows the time consumed by FlowDroid and \tool{}, respectively.
On average, it takes 177.09 seconds for \tool{} to process an app in our dataset, which is comparable to that of the original FlowDroid (i.e., on average, 132.74 seconds to process an app).
As experimentally demonstrated by Avdiienko et al.~\cite{avdiienko2015mining}, by increasing the capacity of the execution server, the performance of FlowDroid could be further improved.
This improvement should also be applicable to \tool{}, making it also possible to analyze real-world apps in practice.
The fact that the time difference between \tool{} and FlowDroid is relatively small suggests that it is also capable of applying \tool{} to analyze (in parallel) large-scale Android apps, as what has been experimentally demonstrated to be true for FlowDroid.


\begin{tcolorbox}[title=\textbf{RQ3 \ding{43} Efficiency}, left=2pt, right=2pt,top=2pt,bottom=2pt]
The time consumption of \tool{} is acceptable for real-time sensor leak detection, with on average 177.09 seconds for one app  without a high increase when comparing with FlowDroid, which is suitable for real-time app analysis.
\end{tcolorbox}







\section{Discussion}

We now discuss the potential implications and limitations of this work.

\subsection{Implications}

\textbf{Beyond smartphone apps.}
The motivating example presented in Section~\ref{subsec:motivation} is extracted from an attack targeting smartwatches, which also supply sensors to support client apps to implement advanced features.
These sensors could be abused by smartwatch app developers, especially malicious attackers.
We argue there is also a strong need to characterize sensor leaks in smartwatch apps, not just smart phones.
Our preliminary experiment has shown that \tool{} can be directly applied to correctly pinpoint the sensor leaks in the Android-based smartwatch apps that sniff passwords~\cite{chen2021comparative}.

Android has been used on more and more devices, such as TVs, home appliances, fitness machines and cars.
The apps in these devices could also all be compromised to leak end-users' sensitive data and hence should also be carefully analyzed before releasing them to the public.
\tool{} could also be useful to characterize data leaks for such Android devices and we will examine some of these in our future work.

\textbf{Beyond sensor leaks.}
As argued by Zhang et al.~\cite{zhangcondysta}, tainted values of string type could be organized as fields in objects, which cannot be detected by state-of-the-art static taint analysis tools such as  FlowDroid. This is because FlowDroid only supports methods as sources. Thus, sensitive field sources are overlooked by FlowDroid, giving rise to many false negatives. 

Our \tool{} extends FlowDroid to mitigate this research gap by introducing field-triggered static taint analysis. 
It is worth highlighting that \tool{} is capable of not only detecting sensor leaks but also pinpointing general privacy leaks, either triggered by source methods or fields.
To help users experience this feature, we have committed a pull request to the original FlowDroid on GitHub so that users can easily access Field-triggered Static Taint Analysis by simply configuring their interested field sources in \emph{SourcesAndSinks.txt} file. 

\textbf{Automated approaches for discovering sensitive source fields.}
In this work, the sensor-related sensitive source fields are identified through manual effort. These are well known to be time-intensive and error-prone.
Hence, our current \tool{} approach is not directly applicable for detecting general field-triggered privacy leaks.
To achieve this, we need to go through all the fields defined in the Android framework to identify sensitive ones. This is non-trivial as Android is now one of the largest community software projects and contains nearly 10K classes.
There is a need to invent new automated approaches to discover sensitive source fields.
One possible solution would be to extend the machine learning approach applied in the SUSI tool to support the prediction of sensitive source fields.

\subsection{Limitations of \tool{}}
\label{subsec:limitations}
\textbf{Limitation of static analysis.}
One major limitation of our tool lies in the intrinsic vulnerability of static code analysis when encountering code obfuscation, reflection, native code, etc. These lead to the unsoundness of our approach. However, these challenges are regarded as well known and non-trivial issues to overcome in our research community.  In our future work, we want to integrate other useful tools developed by our fellow researchers to overcome these shortcomings. For example, we plan to leverage DrodRA \cite{sun2021taming, li2016droidra} to reduce the impact of reflective calls on our static analysis approach.

As explained in Section \ref{subsec:type}, our sensor type inference approach can not trace the sensor type in method-triggered leaks when multiple sensors are available on a device. This is because the actual calling object of a method can only be obtained at run-time. We plan to overcome this limitation in our future work by incorporating dynamic analysis approaches to obtain the required run-time values.

\textbf{Limitations inherited from FlowDroid.}
Since our \tool{} approach directly extends FlowDroid to detect sensor-triggered privacy leaks, it also has all of the limitations of FlowDroid.
For example,  FlowDroid may yield unsound results because it may have overlooked certain callback methods involved in the Android lifecycle or incorrectly modelled native methods accessed by the app.
FlowDroid is also oblivious to multi-threading and it assumes threads to execute in an arbitrary but sequential order, which may also lead to false results.

\textbf{Limitations inherited from SUSI.}
The sensor-related sensitive source methods are collected based on the results of the state-of-the-art tool SUSI. This is also the tool leveraged by the FlowDroid to identify source and sink methods.
However, the results of SUSI may not be  completely correct -- some of its identified sources may not be truly sensitive.
However, this threat has no impact on our approach but only on our experimental results.
This limitation could be mitigated if a better set of source and sink methods are configured.

\textbf{Threats to Validity.}
Apart from these technical limitations, our work also involves some manual efforts. For example, the sensor-related sensitive source fields are summarized manually by reading the Android developers' documentation.
Such manual processes may also introduce errors of their own.
To mitigate this threat, the authors of this paper have cross-validated the results, and we release our tool\footnote{https://github.com/MobileSE/SEEKER} and dataset\footnote{https://zenodo.org/record/4764311\#.YJ91jJMzadZ} for public access.

\section{RELATED WORK}

\textbf{Android sensor usage.} Android sensor usage has long been analyzed in software security mechanisms. Related works \cite{zhu2013sensec, ba2020learning,xu2012taplogger,miluzzo2012tapprints,liu2015exploring,aviv2012practicality,cai2011touchlogger,owusu2012accessory,lee2015multi} have indicated that embedded sensors can be intentionally misused by malicious apps for privacy compromise. Ba et al. \cite{ba2020learning} proposed a side-channel attack that adopts accelerometer data to eavesdrop on the speaker in smartphones. Xu et al. \cite{xu2012taplogger} have shown that it is feasible to infer user's tap inputs using its integrated motion sensors. Liang Cai et al.\cite{cai2011touchlogger} revealed that confidential data could be leaked when motion sensors, such as accelerometers and gyroscopes, are used to infer keystrokes.
Also, Lin et al.\cite{lin2012new} demonstrated that the orientation sensor of the smartphone could be utilized to detect users' unique gesture to hold and operate their smartphones.

Android Sensor misuse is one of the major causes of privacy leaks and security issues on the Android platform. Zhu et al. \cite{zhu2013sensec} collected sensor data from accelerometers, gyroscopes and magnetometers and constructs users' gesture based on these data. Their work indicates that it is feasible to get access to sensory data for personalized usage. Liu et al. \cite{liu2015exploring} demonstrated the most frequently used sensors in Android devices and revealed their usage patterns through backward tracking analysis. They further investigate sensor data propagation path for accurately characterizing the sensor usage \cite{liu2018discovering}. Their findings suggest that the accelerometer is the most frequently used sensor  and the sensor data are always used in local codes.

\textbf{Software side-channels attacks.}
Many previous studies \cite{chang2009inferring, lester2004you, liu2009uwave, ravi2005activity, allen2006classification, schlegel2011soundcomber} explored password inference through specific sensors on smartphones. Owusu et al. \cite{owusu2012accessory} showed that accelerometer values could be used as a powerful side channel to figure out the password on a touchscreen keyboard. Cai et al.\cite{cai2011touchlogger} provided insights of how motion sensors, such as accelerometers and
gyroscopes, can be used to infer keystrokes. Cai et al. \cite{cai2009defending} found that mobile phone sensors are inadequately protected by permission system so that it can raise serious privacy concerns. Enck et al. \cite{enck2014taintdroid} developed TaintDroid that takes sensor information (i.e., location and accelerometer) as sources to detect privacy leaks. Mehrnezhad et al. \cite{mehrnezhad2018stealing} show that orientation sensor can be stealthily listened to without requesting any permission, contributing for attackers to infer the user’s PIN. However, these works emphasize the challenges facing the detection of sensor-sniffing apps or only provided specific attacks by using sensor data. None of them can systematically characterize data leaks in all kinds of sensors.

\textbf{Static analysis on Android apps.}
Android users have long been suffered from privacy leaks \cite{li2017static, kong2018automated, samhi2021raicc, octeau2016combining}. Several solutions have been proposed for detecting such data leaks through static taint analysis~\cite{gao2020borrowing, li2015apkcombiner, yang2017characterizing}. For example, Arzt et al. \cite{arzt2014flowdroid} developed FlowDroid, a context, flow, field and object-sensitive static analysis tool for detecting potential data leaks in Android Apps. Based on Soot \cite{vallee2010soot}, FlowDroid relies on pre-defined knowledge to pinpoint taint flows between source and sink APIs. Zhang et al. \cite{zhangcondysta} developed ConDySTA, a dynamic taint analysis approach, as a supplement to static taint analysis by introducing inaccessible code and sources that help reduce false negatives. Further, Li et al. \cite{li2015iccta} presented IccTA, which can precisely perform data-flow analysis across multiple components for Android apps. Klieber et al. \cite{klieber2014android} augment the FlowDroid and Epicc\cite{octeau2013effective} analyses by tracking both inter-component and intra-component data flow in Android apps.
However, none of these tools concerns the leaks that originated from sensors. Apart from that, our tool only takes the sensor-related code into account, which cost less time by pruning the control flow graph.

The most similar work to ours is SDFDroid\cite{liu2018discovering}, which provides the sensor usage patterns through data flow analysis. As a static approach, however, it focuses on different research object compared to \tool{}. For example, SDFDroid reveals sensor usage patterns while our work explores how and where the sensor data are leaked. On the other hand, SDFDroid applies a static approach to extract sensor data propagation path to construct sensor usage patterns through clustering analysis. In contrast to SDFDroid, \tool{} provide detailed privacy leaks caused by misuse of sensor data, which haven't been found by SDFDroid.

\section{CONCLUSION}
We have presented a novel tool, \tool{}, for characterizing sensor leaks in Android apps. Our experimental results on a large scale of real-world Android apps indicate that our tool is effective in identifying all types of potential sensor leaks in Android apps. Our tool is not only capable of detecting sensor leaks, but also pinpointing general privacy leaks that are triggered by class fields. 
Although there are related works on sensor usage analysis, to the best of our knowledge, there is no other work that thoroughly analyses Android sensor leakage. Unlike previous works, our tool is the first one  to characterize all kinds of sensor leaks in Android apps. We extend FlowDroid for supporting field sources detection (i.e., merged to FlowDroid via pull \#385 on Github\cite{FlowDroidMerge}), which we believe could be adapted to analyze other sensitive field-triggered leaks.
To benefit our fellow researchers and practitioners towards achieving this, we have made our approach open source at the following Github site.

\begin{center}
   \url{https://github.com/MobileSE/SEEKER} 
\end{center}

\section*{Acknowledgements}
{
This work was supported by the Australian Research Council (ARC) under a Laureate Fellowship project FL190100035, a Discovery Early Career Researcher Award (DECRA) project DE200100016, and a Discovery project DP200100020.
This research was also supported by the Open Project Program of the State Key Laboratory of Mathematical Engineering and Advanced Computing (No. 2020A06) and the Open Project Program of the Key Laboratory of Safety-Critical Software (NUAA), Ministry of Industry and Information Technology (No. XCA20026).
}

\balance
\bibliographystyle{IEEEtran}
\bibliography{ref}

\end{document}